# Zn-induced in-gap electronic states in La214 probed by uniform magnetic susceptibility: relevance to the suppression of superconducting $T_c$

R. S. Islam and S. H. Naqib[*]

*Department of Physics, University of Rajshahi, Rajshahi-6205, Bangladesh*

[*]Corresponding author. Email: salehnaqib@yahoo.com

**Abstract**

Substitution of isovalent non-magnetic defects, such as Zn, in $CuO_2$ plane strongly modifies the magnetic properties of strongly electron correlated hole doped cuprate superconductors. The reason for enhanced uniform magnetic susceptibility, $\chi$, in Zn substituted cuprates is debatable. So far, the observed magnetic behavior has been analyzed mainly in terms of two somewhat contrasting scenarios, (a) that due to independent localized moments appearing in the vicinity of Zn arising because of the strong electronic/magnetic correlations present in the host compound and (b) that due to transfer of quasiparticle spectral weight and creation of weakly localized low energy electronic states associated with each Zn atom in place of an in-plane Cu. If the second scenario is correct, one should expect a direct correspondence between Zn induced suppression of superconducting transition temperature, $T_c$, and the extent of the enhanced magnetic susceptibility at low temperature. In this case, the low-$T$ enhancement of $\chi$ would be due to weakly localized quasiparticle states at low energy and these electronic states will be precluded from taking part in Cooper pairing. We explore this second possibility by analyzing the $\chi(T)$ data for $La_{2-x}Sr_xCu_{1-y}Zn_yO_4$ with different hole contents, $p$ ($= x$), and Zn concentrations ($y$) in this paper. Results of our analysis support this scenario.

**Keywords**: High-$T_c$ cuprates; effect of disorder; magnetic susceptibility; pseudogap

**PACS**: 74.72.Dn, 74.25.Ha, 74.25.-Dw, 74.62.Dh

# 1 Introduction

The observation that isovalent non-magnetic defects in $CuO_2$ planes of hole doped cuprates reduces the superconducting $T_c$ most effectively [1], triggered a surge of theoretical and experimental activities since the early days of their discovery. Besides, the low-energy quasiparticle (QP) dynamics in strongly electronically correlated cuprates can be probed very efficiently via in-plane defect studies [2]. Today this strong suppression of $T_c$ is understood within the framework of disorder induced potential (unitary) scattering of Cooper pairs with *d*-wave order parameter [3]. The second striking effect of these non-magnetic defects is the impurity-induced magnetism. EPR [4], NMR [5], and bulk magnetic susceptibility [6] experiments gave some indications that isovalent non-magnetic defects like Zn give rise to a local magnetic moment-like feature on the nearest neighbors of the substituted Cu sites in the $CuO_2$ planes. This apparent magnetic behavior manifests itself as a Curie-like term in the bulk temperature dependent magnetic susceptibility, $\chi(T)$ [6] at low temperatures. Some researchers have attributed this effect to the uncompensated spins (spin vacancy) in the presence of short-range antiferromagnetic (AF) correlations, arising as a result of replacing ($s = 1/2$) $Cu^{2+}$ with isovalent and non-magnetic $Zn^{2+}$ ($s = 0$). On the other hand, pioneering scanning tunneling microscopy (STM) experiments on Zn-substituted hole doped cuprates by Pan *et al.* [7] have found intense low-energy QP scattering resonances at the Zn sites, coincident with the strong suppression of superconductivity within a length scale ~ 15 Å around the defect sites.

The observed magnetic behavior of impurity substituted hole doped cuprates has been modeled mainly by using two different frameworks. Both these frameworks describe experimental results with some degree of success but the underlying physics is markedly different. The first scenario invokes a picture where the appearance of localized spins in the neighboring Cu sites at low temperatures enhances the uniform static magnetic susceptibility, whereas the second scenario implies that the magnetic effect is basically due to weakly localized QP spectral weight, i.e., an effect arising from the electronic energy density of states (EDOS) at low-energy [8, 9]. It is fair to say that the situation remains unsettled till date [8 – 10].

In this work we wish to address this issue by looking at the Zn induced suppression of superconducting transition temperature in hole doped cuprates. We assume that the weakly localized low-energy quasiparticles created due to Zn substitution do not take part in Cooper pairing. We also assume that, it is these quasiparticle states which induce a low-$T$ Curie-like contribution in the bulk magnetic susceptibility. Based on these assumptions, we have analyzed the $T_c$ suppression data and the $\chi(T)$ data for La$_{2-x}$Sr$_x$Cu$_{1-y}$Zn$_y$O$_4$ with different compositions ($x$ and $y$ values). A consistent picture emerges which supports the scenario where Zn induced modifications of $\chi(T)$ originates from the underlying changes in the low-energy EDOS.

The rest of the paper is organized as follows. We present the experimental data in Section 2. Analysis of data is presented in Section 3. Results of analysis of are discussed in Section 4. Finally, Section 5 consists of the concluding remarks.

## 2 Experimental samples and experimental results

Polycrystalline single-phase sintered samples of La$_{2-x}$Sr$_x$Cu$_{1-y}$Zn$_y$O$_4$ were synthesized by standard solid-state reaction methods using high-purity (> 99.99%) starting materials. Samples were characterized by X-ray powder diffraction, room-temperature thermopower, and AC magnetic susceptibility (ACS) measurements. Details of sample preparation and characterization can be found in Refs. 11 and 12. $\chi(T)$ measurements were done using a model 1822 *Quantum Design* SQUID magnetometer. Data were collected following a predefined sequence, usually in the range of 5 K to 400 K, under an applied magnetic field of 5 Tesla. The data were corrected for background and molar bulk magnetic susceptibility was calculated. In this study we have used two different Sr contents – $x$ = 0.09 (underdoped) and 0.15 (close to the optimum doping), each with the following nominal Zn concentrations – $y$ = 0.0, 0.005, 0.01, 0.015, 0.02, and 0.024.

Illustrative X-ray powder diffraction profiles are shown in Figs. 1. Copper $K_{\alpha1}$ radiation was used (wavelength 1.541Å). Within the resolution of the X-ray diffractometer, the samples were found to be almost phase pure. In a few of the compounds impurity phases (mainly LaCuO$_3$) with an intensity less than 1% was detected. The superconducting transition temperature, $T_c$, was measured from low-field AC susceptibility data. The $T_c$ values with different hole and Zn contents are tabulated in Table 1. $T_c$ values are accurate within ± 0.5 K. The variation of $T_c$ with Zn content for different hole concentrations is shown in Fig. 2. The room-temperature thermopower is sensitive to the hole content and therefore, to the Sr content in LSCO [13]. Thermopower data reveals that Zn does not change the hole concentration in the $CuO_2$ plane, as found in other studies [14 – 17]. The background signal corrected $\chi(T)$ data are shown in Figs. 3. Inspection of these data reveals that a Curie-like enhancement appears only at low-$T$ in the uniform magnetic susceptibility and increases systematically with Zn content. The increment is markedly stronger in the underdoped compounds. At this point, it is worth mentioning that, unlike in the case of magnetic Ni or Co substituted high-$T_c$ compounds [18], the Curie-like increase in the magnetic susceptibility due to non-magnetic Zn appears only at relatively low-$T$ region. There is also a gradual suppression of high-$T$ $\chi(T)$. These two effects together can lead to a crossing point in $\chi(T)$ which shifts to lower $T$ with increasing hole content (see Figs. 3). It is widely believed that, in some way or other, the low-$T$ enhancement almost surely arises due to the strong electronic correlations and the pseudogap present in the host material [8, 10, 19].

**3 Analysis of experimental data**

(a) *Suppression of $T_c$ due to Zn*

It is seen from Fig. 2 that irrespective of hole content, Zn suppresses $T_c$ almost linearly. The rate of suppression $dT_c/dy$, is significantly higher for the underdoped compound ($dT_c/dy$ = -12.1/%Zn and -16.2/%Zn for $x$ = 0.15 and $x$ = 0.09 compounds, respectively). This hole content dependent value of $dT_c/dy$ can be accounted for by considering a Zn induced pair-breaking picture in the unitary limit [3, 20, 21]. In this scenario, each non-magnetic Zn roughly breaks a $d$-wave Cooper pair where the pair-breaking scattering rate varies inversely with the EDOS around the Fermi level. The usual pair-breaking scheme can be expressed as [22]-

$$-\ln\left(\frac{T_c}{T_{c0}}\right) = \Psi\left[\frac{1}{2} + \frac{\Gamma}{2\pi k_B T_c}\right] - \Psi\left[\frac{1}{2}\right] \qquad (1)$$

Here, $\Psi$ is the digamma function, $T_{c0} = T_c$ ($y$ = 0), and $\Gamma$ is the pair-breaking scattering rate (in energy unit), for potential scattering. This scattering rate can be expressed as $\Gamma = y/\pi N(\varepsilon_F)$. $N(\varepsilon_F)$ should be taken as the thermally averaged (over a temperature window of width ~ ± $2k_BT_c$ about the Fermi energy) EDOS around the Fermi level. Assuming a triangular pseudogap (PG) [18, 19, 23, 24], one can calculate the pair-breaking scattering rate within this framework. The pair-breaking scattering rate is found to vary as $T^*/T_c$ for a fixed impurity concentration, $y$. $T^*$ is the characteristic PG temperature, related roughly to the characteristic PG energy scale, $\varepsilon_g \equiv k_B T^*$. This scheme, therefore, provides one with a natural explanation for the hole content dependent values of $dT_c/dy$ in terms of the PG in the QP spectral density. PG increases with underdoping and thereby reduces the EDOS at the Fermi level, which in turn increases the pair-breaking scattering rate and $T_c$ is reduced more rapidly with impurity content as a consequence. For deeply overdoped compounds, where PG is absent, the scattering rate should become nearly constant and $dT_c/dy$ should become almost hole content independent, as is seen in experiments [3, 20, 21].

We have shown the hole content dependent rate of suppression of $T_c$ and the corresponding PG energy scales for $La_{2-x}Sr_xCu_{1-y}Zn_yO_4$ and $Y_{1-x}Ca_xBa_2(Cu_{1-y}Zn_y)_3O_{7-\delta}$ [8, 15 – 17, 21, 24, 25] in Fig. 4. Fig. 4 shows a clear anti-correlation between $dT_c(p)/dy$ and $\varepsilon_g(p)$. This is a generic feature followed by all the different families of hole doped cuprates [3, 20, 26].

(b) *Effect of Zn on uniform magnetic susceptibility*

Figs. 3 shows the experimental $\chi(T, p)$ data of the compounds under study. In previous studies [9, 11, 12, 24, 27] we have demonstrated that the $\chi(T, p)$ behavior of different hole doped cuprate families follows mainly from the $p$ and $T$ dependent Pauli spin susceptibility ($\chi_s$) part. The $p$ and $T$ dependent features of $\chi(T, p)$ primarily follow from the $p$ dependent PG energy scale [9, 11, 12, 24, 27, 28]. Strong experimental support is found from the analysis of specific heat and NMR data [29 - 32]. In many ways, results obtained from $\chi(T, p)$ complement those obtained from the pioneering specific heat studies by Loram *et al.* [18, 23, 30]. One of the most unambiguous measures of QP spectral density is obtained from the electronic specific heat coefficient, $\gamma$. The striking qualitative and quantitative resemblance between bulk magnetic susceptibility and electronic specific heat has been well documented [18, 23, 30]. $\chi_s(T, p)$ and $S(T, p)/T$ (where, the electronic entropy $S(T, p) = \int \gamma(T, p)dT$) show completely identical $T$ dependences. Furthermore, the value of $S/\chi_s T$ is very close to the free electron Wilson ratio $a_0 \equiv \pi^2 k_B^2/3\mu_B^2$. The similar $T$-dependences of $S/T$ and $\chi_s$ show that the total spin and charge excitation spectrum have similar energy dependences. The fact that $S/\chi_s T \sim a_0$, does not require that the respective QP spectra are those of free electrons, but rather that the relative number of thermally excited charge and spin excitations is that expected for conventional Fermions in cuprates, even in the presence of strong electronic correlations in these materials.

Based on the above experimental and theoretical indications and our previous analysis of the $\chi(T, p)$ data for different families of cuprates, including LSCO [9, 11, 12, 24, 27], we propose a simple model to analyze the Zn induced behavior of $\chi(T, p)$ as follows. We assume that the quasiparticles within the resonance peak appear at the expense of the states on the flanks of the pseudogap, thereby expanding the pseudogap locally (see Fig. 5). This takes into account the experimental observation that $\chi(T)T$ for different Zn concentration at a given hole content, are not parallel at high-$T$ (the apparent 'melting' of Zn induced 'magnetic moment' at high-$T$) [8, 11, 12] for underdoped compounds. From this it can be inferred that there are no extra states inside the PG region in the EDOS, only a rearrangement of electronic states take place. An approximated resonance bar of width $2\varepsilon$ can be employed to fit the uniform magnetic susceptibility of Zn-doped LSCO. One can also approximate the QP resonance with a Lorentzian profile [12]. A resonance bar or a Lorentzian yields almost identical qualitative and quantitative results. Following our earlier analyses [11, 12, 24], we have used the following expression to model the magnetic uniform magnetic susceptibility of Zn-doped LSCO,

$$\chi^{PG-Zn}(T) = N_0\mu_B^2[1 - \frac{2k_BT}{\varepsilon_g}\{\ln\cosh(\frac{\varepsilon_g}{2k_BT})\}] + N_1\mu_B^2\tanh(\frac{\varepsilon}{2k_BT}) \qquad (2)$$

In this equation, $N_0$ is related to the electronic density of states of the host LSCO outside the PG region. $N_1$ is the amplitude of the QP resonance density of states.

Before proceeding any further, a few points must be clarified here. Eqn. (2) models the experimental $\chi(T)$ data in terms of thermally averaged EDOS alone. The Zn-induced enhancement of $\chi(T)$ is due to the low-energy QP spectral weight in this scenario and therefore, related to the Pauli paramagnetic effect. But the experimentally measured $\chi(T)$ generally includes several contributions to the bulk magnetic susceptibility that are not directly related to the underlying EDOS, e.g., the core (Larmor) and the Van Vleck contributions. But these contributions are believed to be temperature independent. Landau diamagnetic term on the other hand, has the same energy dependence as the Pauli spin susceptibility. But for hole doped cuprates, this term is only 2% – 5% of the Pauli paramagnetic term in magnitude [33 – 35]. We therefore, do not need to deal with this separately. Thus, in Eqn. (2), $N_0$ not only gives a measure of the flat EDOS outside the PG region, but its value also includes the *T*-independent contributions to the bulk magnetic susceptibility. Results of representative fits to the $\chi_s(T, p)$ data are shown in Fig. 6. The data in the vicinity of the SC transition temperature were not used in the fits. This is because the formalism developed does not take into account the superconducting fluctuations. There is reason to believe that strong pairing fluctuations contributing to fluctuation diamagnetism do not exist 20 – 30 K above $T_c$ in La214 (~ at $T > 1.7\ T_c$) [36 – 38]. The simple model expressed via Eqn. (2) fits the experimental magnetic susceptibility quite satisfactorily. We have shown the fitting parameters in Table 2.

**4 Discussions**

So far, it has been established quite firmly from experimental findings, that the Zn induced magnetic effect is strongly hole content dependent. At the same time the rate of suppression of superconducting $T_c$ is also strongly hole content dependent. Both these observations can be linked to the evolution of the PG in the QP energy density of states with the hole content. A depleted EDOS due to PG enhances the pair-breaking scattering rate due to impurity and $T_c$ is degraded rapidly. Whether there was a direct link between the magnetic enhancement and $T_c$ degradation due to Zn remained as an open question. To investigate this further, we have shown the $\Delta T_c(p, y)$ and the area under the QP resonance, $A(p, y)$, in Fig. 7. Fig. 7 reveals that when hole content is low (PG is large), $T_c$ decreases sharply with in-plane impurity content and at the same time the area under the resonance bar also increases sharply with impurity concentration. This, we believe, is a notable evidence in favor of the scenario where the low-energy QP states are thought to form from a re-distribution of QP states from higher energy to low-energy and these low-energy QPs are excluded from the Cooper pairing condensate. This also supports the previous proposal [8, 11, 12] that Zn induced magnetic effect in hole doped cuprates arises from the redistribution of QP spectral weight rather than appearance of localized Curie moments in the canonical sense. At this point it should be mentioned that Kruis *et al.* [19] proposed a simple model where the presence of a PG ensures a low-energy resonance in the QP spectral density. A triangular PG pinned at the Fermi level was considered in Ref. 19, similar to what we have assumed in this study. Another direct consequence of the proposed scenario is that it offers a natural explanation for the crossing point of the $\chi(T)$ data (Figs. 3a and 3b). As the weakly localized low-energy resonance appears at the expense of the states near the edges of the PG, it leads to a decrease in the $\chi(T)$ at high temperatures (depending on the energy width of the PG) and a subsequent increase in the $\chi(T)$ in the low temperature. These effects result quite naturally in a crossing point in the magnetic susceptibility data as impurity contents are changed. A spin dilution effect also plays a role [8]. This can be traced from a gradually decreasing trend of $N_0$ with impurity concentration (Table 2). The crossing temperature seems to follow the characteristic PG energy scale. Circumstantial evidence also follows from our previous work where we have investigated $\chi(T)$ of overdoped $La_{2-x}Sr_xCu_{1-y}Zn_yO_4$. No crossing point in $\chi(T)$ could be detected [11] in these compounds. This, we believe, is due to the absence of PG in those compounds. In this situation, as suggested by Kruis *et al*. [19], the QP resonant state is formed only when the phase coherent superconducting gap is significant (i.e., below $T_c$). Therefore, its effect in the normal state is absent.

Fig. 7 exhibits the main finding of this study. It is seen clearly that the suppression of $T_c$ ($\Delta T_c(y)$ = $[T_c (y = 0) - T_c (y)]$) with Zn follows closely the growth of the resonance spectral density with Zn. In this work, we suggest that this spectral weight is removed from the superconducting condensate and thereby reduce superfluid density. This proposal qualitatively supports the well-known Uemura relation [39, 40] between superconducting transition temperature and superfluid density. It is also consistent with the *swiss cheese* model on the qualitative level [41]. It is worth mentioning that even though $A(p, y)$ grows linearly with Zn, the extrapolated value of $A(p, y)$ at $y$ = 0 is non vanishing for either of the Sr contents considered in this study. A small residual $A(p, y$ = 0) exists. This perhaps is due to an intrinsic (related essentially to Sr substitution induced disorder) effect or an extrinsic (related to minute level of impurity phase) effect, showing that some disorder capable of generating low energy resonant states exist in the $CuO_2$ plane even when Zn content is zero.

The fits to the $\chi(T)$ data give estimates of the pseudogap energy scale. The extracted values of $\varepsilon_g$ obtained from analysis of uniform magnetic susceptibility agree very well with those extracted from the analysis of heat capacity [18, 23, 29, 30] and charge transport [15 – 17, 42, 43] data of different families of high-$T_c$ cuprates. This gives credence to the model employed in this study. It is important to note that even though Zn acts as an efficient suppressor of $T_c$, it has no effect on $\varepsilon_g$ for a given hole content, within the experimental uncertainty. This contrasting effects of Zn on $T_c$ and $\varepsilon_g$ indicate that Cooper pairing correlations and PG correlations are perhaps not directly linked to each other. A number of previous experimental studies support this proposition [9, 11, 12, 15 – 18, 23, 29, 30, 44]. It is also worth noting that, for a fixed value of $p$, the values of $\varepsilon_g$ are almost independent of the Zn content. For example, the values of $\varepsilon_g$ are $530 \pm 20$ K and $245 \pm 30$ K for $La_{1.91}Sr_xCu_{1-y}Zn_yO_4$ and $La_{1.85}Sr_xCu_{1-y}Zn_yO_4$, respectively. The small and non-systematic variation in the values of $\varepsilon_g$ with Zn for each Sr concentration can be accounted for by an uncertainty in the Sr content at the level of $\Delta x = \pm 0.005$.

In the proposed scenario, the impurity induced resonant QP spectral weight, which is excluded from pairing condensate, contributes to the low-$T$ $\chi(T)$ of $La_{2-x}Sr_xCu_{1-y}Zn_yO_4$. This is consistent with the STM results presented in Ref. 7.

## 4 Conclusions

The effect of Zn on the SC transition temperature and the bulk magnetic susceptibility has been investigated in this paper. An attempt has been made to explain the strong hole content dependences of $dT_c/dy$ and enhanced $\chi(T)$ of $La_{2-x}Sr_xCu_{1-y}Zn_yO_4$ in terms of a model EDOS whose features are dominated by the hole content dependent PG and an impurity induced low-energy QP resonant spectral density. This model yields values of $\varepsilon_g$ in excellent agreement with those found in earlier studies [9, 11, 12, 18, 23, 26, 27, 29, 30]. The model also offers a natural explanation of the crossing of $\chi(T, y)$ data for underdoped cuprates in terms of spin dilution and redistribution of QP spectral weight. The extracted values of $\varepsilon_g$ at a given hole content are Zn independent as found in previous studies [9, 11, 15 – 17, 29]. We have found strong indications that a common mechanism works behind suppression of $T_c$ and enhancement of $\chi(T)$ with Zn. Fig. 7, a central result of this study, implies a direct correspondence between the impurity resonant spectral weight and suppression of $T_c$ in $La_{2-x}Sr_xCu_{1-y}Zn_yO_4$. This provides with strong evidence in favour of the scenario where Zn-induced low-energy resonant states are precluded from SC pairing but contributes to the low-$T$ enhancement of the $\chi(T)$. Both $dT_c/dy$ and enhancement of $\chi(T)$ increase in magnitude with increasing underdoping. This, we believe, is due to the doping dependent behavior of the PG. Theoretical work by Kruis *et al.* [19], supports this assumption. At both qualitative and quantitative level the analyses presented in this study are strongly supported by the electronic heat capacity studies by Loram and co-workers [18, 23, 29, 30]. Since the coefficient of electronic specific heat, $\gamma$, and NMR Knight shifts directly probe the underlying EDOS, an analysis of these parameters for Zn substituted hole doped cuprates, is expected to yield results similar to this study. The Zn induced low-$T$ enhancement of $\gamma$ or the Knight shift should give a direct measure of the unpaired QP resonant spectral weight.

**Acknowledgements**

The authors acknowledge Professor J. R. Cooper and Dr. J. W. Loram (University of Cambridge, UK), and Professor J. L. Tallon (Victoria University, Wellington, New Zealand) for enlightening discussions about effects of disorder in cuprates over the years.

Table 1: Superconducting transition temperatures of La$_{2-x}$Sr$_x$Cu$_{1-y}$Zn$_y$O$_4$ compounds.

| Compound | Zn content, $y$ (%) | $T_c$ (K)* |
|---|---|---|
| La$_{1.91}$Sr$_{0.09}$Cu$_{1-y}$Zn$_y$O$_4$ | 0.0 | 28.46 |
| | 0.5 | 21.40 |
| | 1.0 | 12.65 |
| La$_{1.85}$Sr$_{0.15}$Cu$_{1-y}$Zn$_y$O$_4$ | 0.0 | 36.70 |
| | 0.5 | 30.35 |
| | 1.0 | 25.56 |
| | 1.5 | 19.25 |
| | 2.0 | 11.70 |

* $T_c$ values were obtained from low field (1 Oe) ACS measurements on powdered compounds. $T_c$ values are accurate up to ± 0.5 K.

Table 2: Fitting parameters for fits to the $\chi(T)$ data of La$_{2-x}$Sr$_x$Cu$_{1-y}$Zn$_y$O$_4$ compounds.

| Compound | Zn content, $y$ (%) | $E_g$ (K) | $N_0\mu_B^2$ (10$^{-4}$ emu/mole) | Area under the resonance bar, $A$ (10$^{-4}$ emu-K/mole) |
|---|---|---|---|---|
| La$_{1.91}$Sr$_{0.09}$Cu$_{1-y}$Zn$_y$O$_4$ | 0.0 | 548 | 1.12 | - |
| | 0.5 | 512 | 1.09 | 30.1 |
| | 1.0 | 550 | 1.05 | 41.0 |
| | 2.4 | 539 | 0.97 | 80.1 |
| La$_{1.85}$Sr$_{0.15}$Cu$_{1-y}$Zn$_y$O$_4$ | 0.0 | 258 | 1.06 | - |
| | 0.5 | 275 | 1.03 | 28.8 |
| | 1.0 | 254 | 1.01 | 36.6 |
| | 1.5 | 253 | 0.97 | 42.7 |
| | 2.0 | 217 | 0.92 | 50.1 |
| | 2.4 | 236 | 0.90 | 59.7 |

**Figure captions**

Figure 1(color online): Representative X-ray diffraction profiles of $La_{2-x}Sr_xCu_{1-y}Zn_yO_4$. Sr and Zn contents are given in the plots. Intensities of the diffracted X-ray are shown in square-root (SQRT) scale. The Miller indices and relative intensities of the peaks are also marked.

Figure 2 (color online): $T_c$ $(x, y)$ of $La_{2-x}Sr_xCu_{1-y}Zn_yO_4$. $T_c$ values were obtained from the low-field ACS data. The straight lines are linear least-square fits to the $T_c$ $(y)$ data.

Figure 3: Representative $\chi(T)$ of (a) $La_{1.91}Sr_{0.09}Cu_{1-y}Zn_yO_4$ and (b) $La_{1.85}Sr_{0.15}Cu_{1-y}Zn_yO_4$. Zn contents are given in the plots. The crossing temperature, $T_{cs}$, is also marked (see text for details).

Figure 4 (color online): $dT_c/dy$ (filled symbols, red: $Y_{1-x}Ca_xBa_2(Cu_{1-y}Zn_y)_3O_{7-\delta}$; blue: $La_{2-x}Sr_xCu_{1-y}Zn_yO_4$) and characteristic pseudogap energy(expressed in temperature), $\varepsilon_g$, (empty symbols, red: $Y_{1-x}Ca_xBa_2(Cu_{1-y}Zn_y)_3O_{7-\delta}$; blue: $La_{2-x}Sr_xCu_{1-y}Zn_yO_4$) as a function of in-plane doped hole content. Dashed straight lines are drawn as guides to the eyes. PG energies shown in this plot is for the Zn-free compounds ($\varepsilon_g$-values are independent of the Zn content).

Figure 5 (color online): Model electronic density of states and the impurity resonance. A triangular PG has been assumed. The black lines show the host EDOS. The dotted lines show the EDOS around an impurity. The arrows show the states involved in the formation of the resonant QP states. These states are excluded from pairing condensate (see text for details).

Figure 6 (color online): Representative fits to experimental $\chi(T)$ of (a) $La_{1.91}Sr_{0.09}Cu_{1-y}Zn_yO_4$ and (b) $La_{1.85}Sr_{0.15}Cu_{1-y}Zn_yO_4$. Zn contents are given in the plots. The dashed lines are the fits. To avoid contributions from superconducting fluctuations, low-$T$ data below 50 K have been avoided. For clarity, every one of two experimental data points is shown.

Figure 7 (color online): $\Delta T_c(p, y)$ and the resonance spectral weight ($A(p, y)$) of $La_{1.85}Sr_{0.15}Cu_{1-y}Zn_yO_4$ (main panel) and $La_{1.91}Sr_{0.09}Cu_{1-y}Zn_yO_4$ (inset).

Figure 1

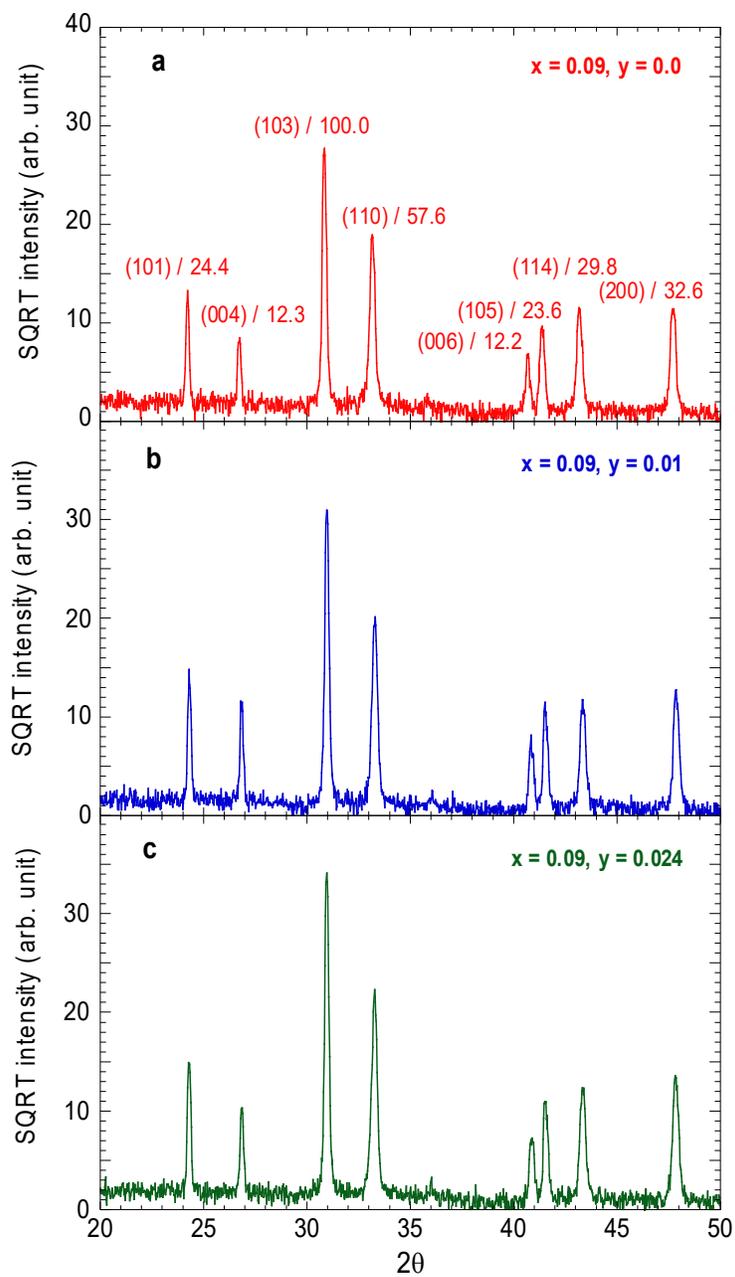

Figure 2

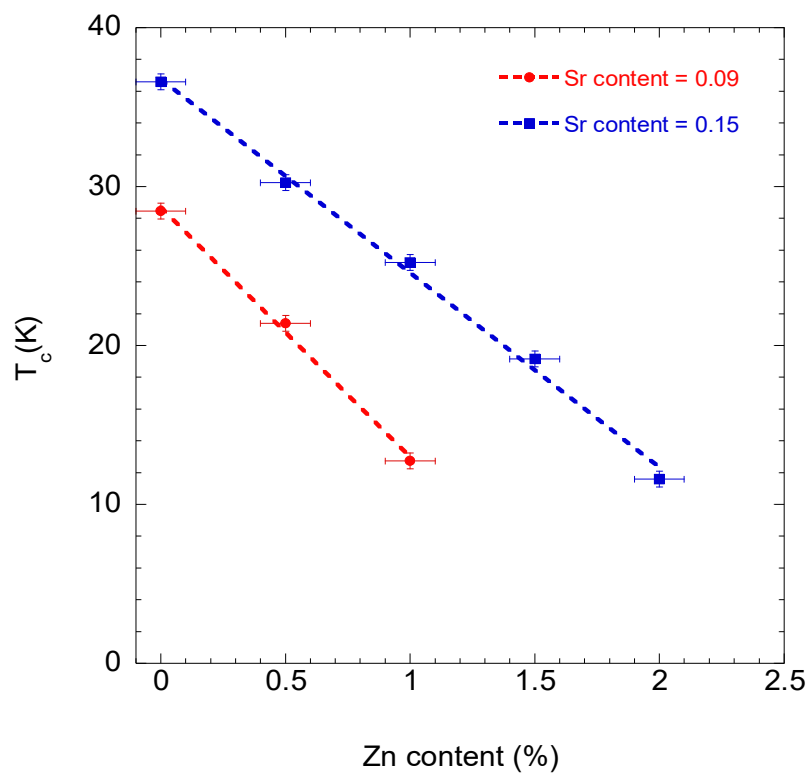

Figure 3

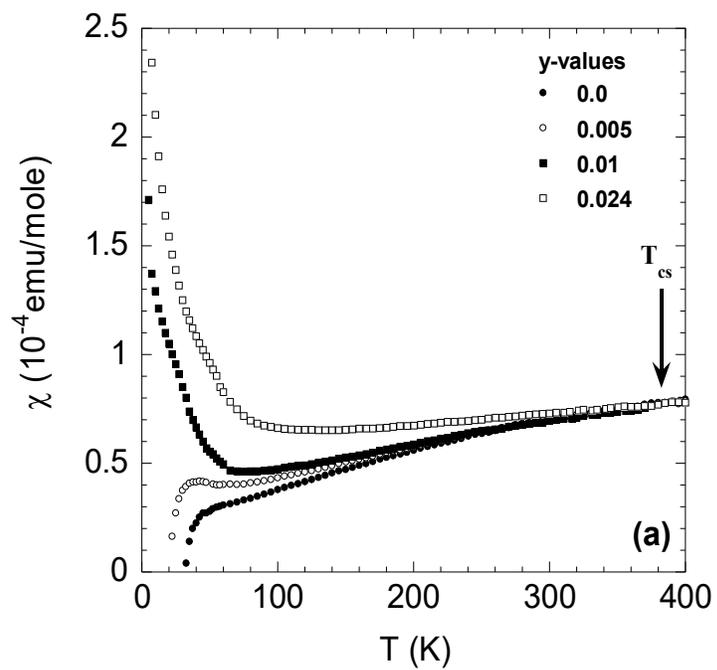

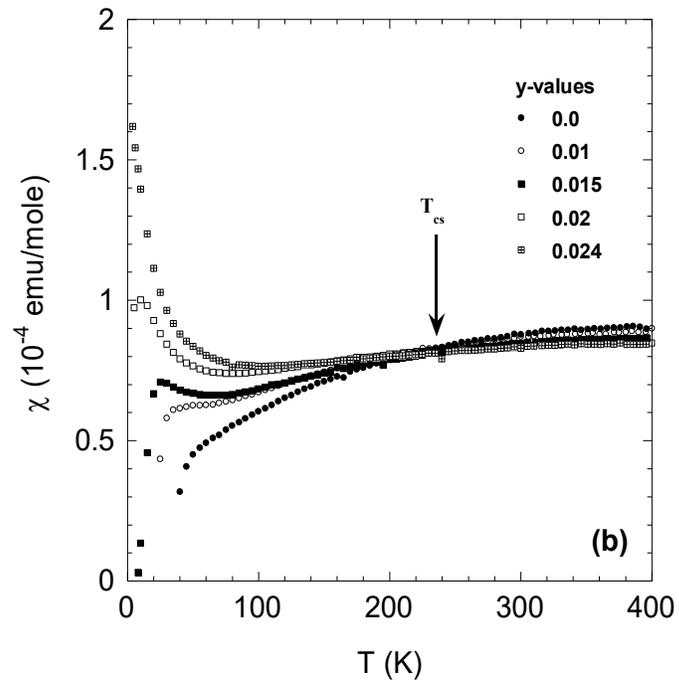

Figure 4

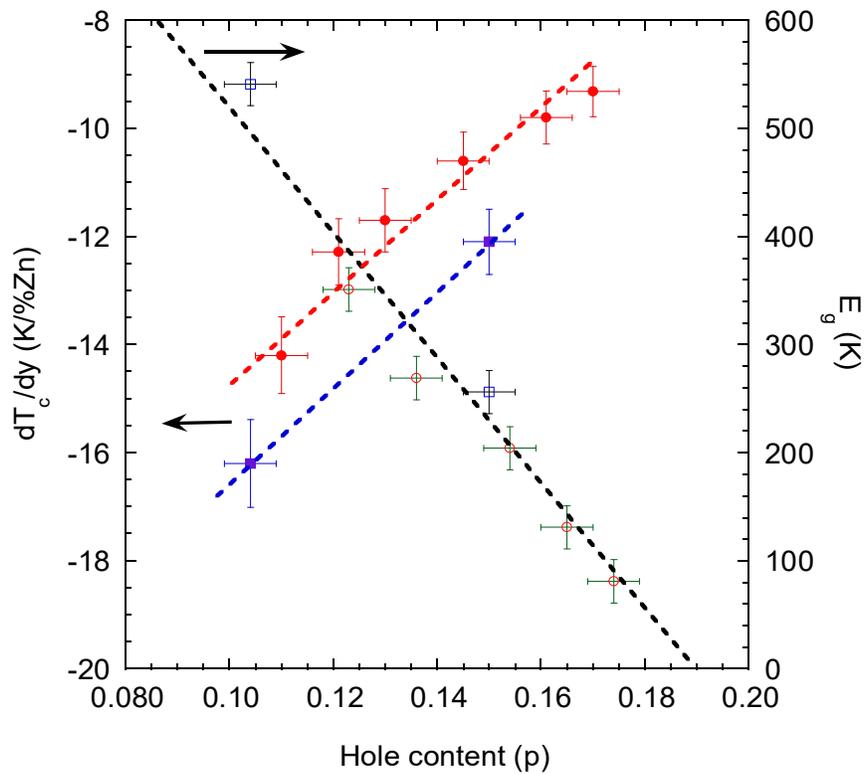

Figure 5

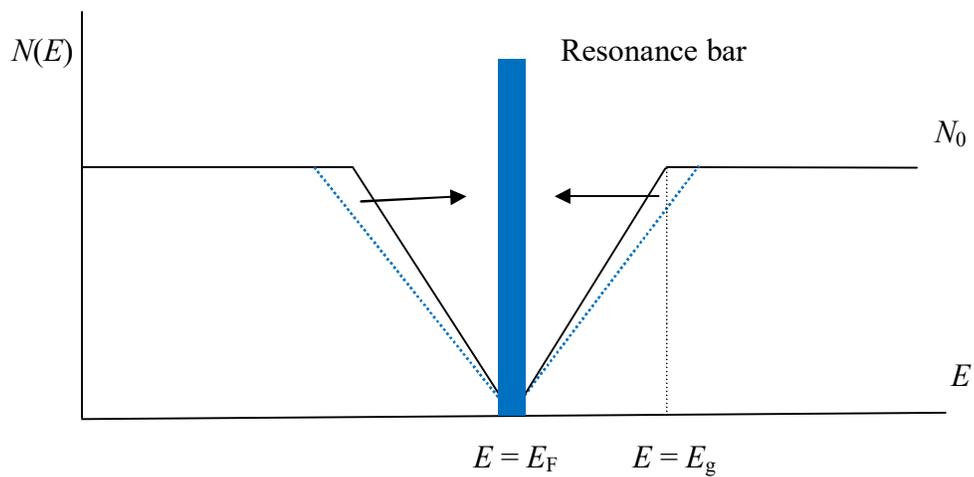

Figure 6

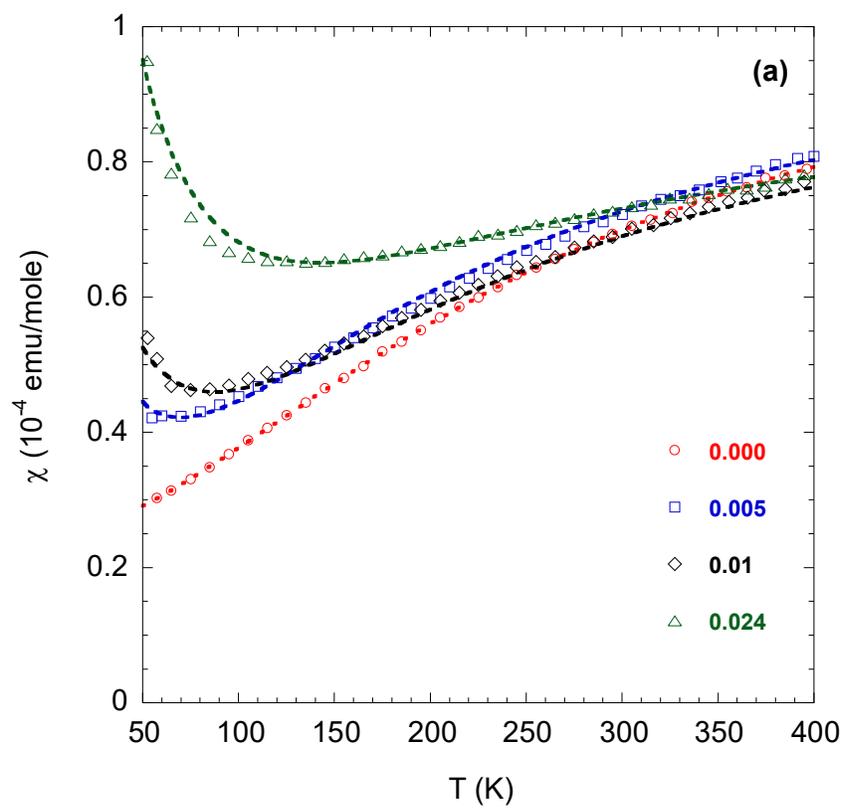

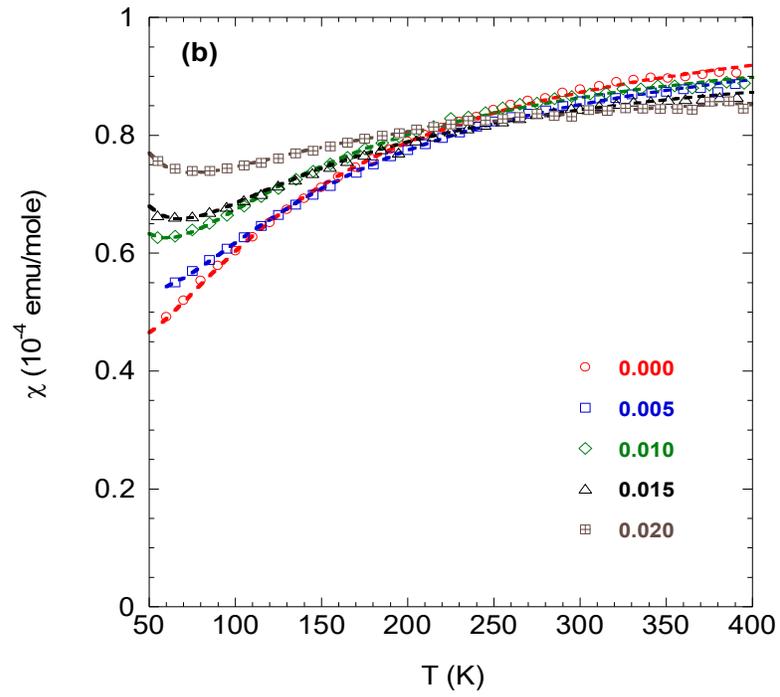

Figure 7

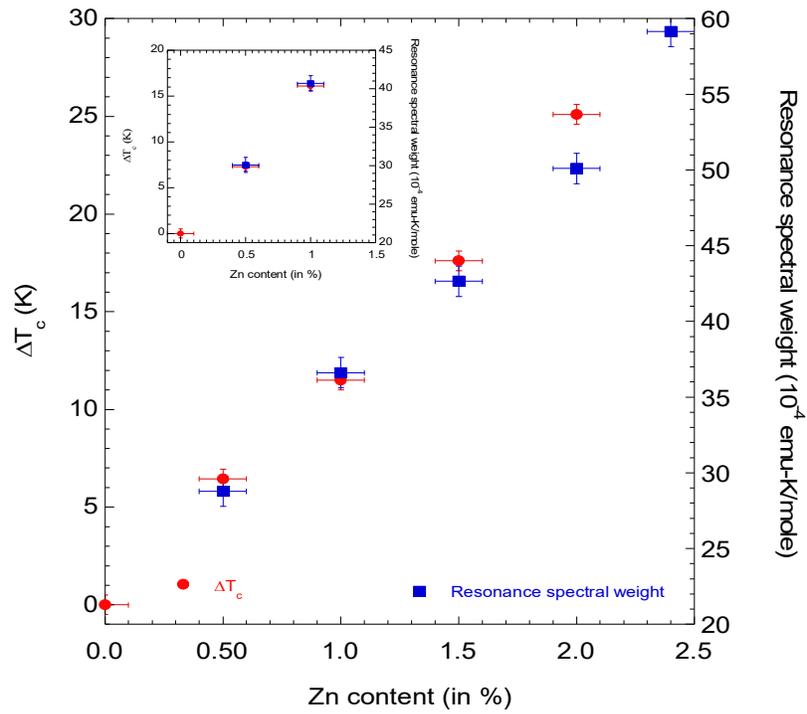